\documentclass[journal]{IEEEtran}
\usepackage{enumerate}
\usepackage{amsmath,amssymb,verbatim,cite,amsopn,graphicx,citesort,url,color}
\newcommand{\vr}{\gamma}

\newcommand{\vrhb}{\hat{\gamma}_b}
\newcommand{\vrtb}{\tilde{\gamma}_b}
\newcommand{\vrhba}{P_b(1-\beta_b)}
\newcommand{\vrtba}{P_b\beta_b}
\newcommand{\vrea}{P_e}
\newcommand{\vrhe}{\hat{\gamma}_e}
\newcommand{\vrte}{\tilde{\gamma}_e}
\newcommand{\vrhea}{P_e(1-\beta_e)}
\newcommand{\vrtea}{P_e\beta_e}
\newtheorem{Proposition}{Proposition}

\begin{document}
\title{Secure On-Off Transmission Design with Channel Estimation Errors}
\author{
Biao~He,~\IEEEmembership{Student~Member,~IEEE,}
and~Xiangyun~Zhou,~\IEEEmembership{Member,~IEEE}
\thanks{B. He and X. Zhou are with the Research School of Engineering, the Australian National University, Australia (e-mail: biao.he@anu.edu.au, xiangyun.zhou@anu.edu.au).}
\thanks{The material in this paper was presented in part at Australian Communications Theory Workshop (AusCTW), Adelaide, Australia, Jan. 2013~\cite{He_13}. This work was supported by the Australian Research Council's Discovery Projects funding scheme (project no. DP110102548).}
}
\maketitle

\begin{abstract}
Physical layer security has recently been regarded as an emerging technique to complement and improve the communication security in future wireless networks. The current research and development in physical layer security is often based on the ideal assumption of perfect channel knowledge or the capability of variable-rate transmissions.
In this work, we study the secure transmission design in more practical scenarios by considering channel estimation errors at the receiver and investigating both fixed-rate and variable-rate transmissions.
Assuming quasi-static fading channels, we design secure on-off transmission schemes to maximize the throughput subject to a constraint on secrecy outage probability.
For systems with given and fixed encoding rates, we show how the optimal on-off transmission thresholds and the achievable throughput vary with the amount of knowledge on the eavesdropper's channel.
In particular, our design covers the interesting case where the eavesdropper also uses the pilots sent from the transmitter to obtain imperfect channel estimation.
An interesting observation is that using too much pilot power can harm the throughput of secure transmission if both the legitimate receiver and the eavesdropper have channel estimation errors, while the secure transmission always benefits from increasing pilot power when only the legitimate receiver has channel estimation errors but not the eavesdropper.
When the encoding rates are controllable parameters to design, we further derive both a non-adaptive and an adaptive rate transmission schemes by jointly optimizing the encoding rates and the on-off transmission thresholds to maximize the throughput of secure transmissions.
\end{abstract}


\section{Introduction} \label{sec:Intro}
\IEEEPARstart{T}{he} broadcast nature of wireless networks makes communication security a critical issue, especially when the information transmitted is important and private.
Cryptographic technologies are traditionally used to increase the wireless communication security.
On the other hand, physical layer security has been widely regarded as a complement to cryptographic technologies in future networks.
Wyner's pioneering work introduced the wiretap channel model as a basic framework for physical layer security~\cite{wyner_75}, which was extended to broadcast channels with confidential messages described by Csisz\'{a}r and K\"{o}rner in~\cite{csiszar_78}.
These early works have led to a significant amount of recent research activities taking the fading characteristics of wireless channels into account.
One of the key features in providing physical layer security is that the channel state information (CSI) of both the legitimate receiver and the eavesdropper often needs to be known by the transmitter to enable secure encoding and advanced signaling.
In recent years, increasing attention has been paid to the impact of the uncertainty in the CSI of both legitimate receiver and eavesdropper's channels at the transmitter, e.g.,~\cite{Mukherjee_11,Ng_11,Li_11,Lin_11,Huang_12}.

Usually, the CSI is obtained at the receiver by channel estimation during pilot transmission. Then, a feedback link (if available) is used to send the CSI to the transmitter. Hence, the accuracy of the channel estimation at the receiver affects the quality of CSI at the transmitter.
In the literature of physical layer security, most existing studies assumed that the legitimate receiver has perfect channel estimation.
Clearly, this assumption is not very practical, since the channel estimation problem generally is not error-free.
In principle, the channel estimation error exists at both the legitimate receiver and the eavesdropper.
Assuming perfect estimation at the eavesdropper is more reasonable from the secure transmission design point of view, since it is often difficult or impossible for the transmitter to know the accuracy of the eavesdropper's channel estimate.
Nevertheless, in scenarios where the eavesdropper is just an ordinary user of the network whose performance and other information can be tracked by the transmitter, e.g.,~\cite{Liang_08,Khisti_10,Leow_11},
the consideration of imperfect channel estimation at the eavesdropper becomes relevant.
Previous works that study the physical layer security problems considering the imperfect channel estimation at the receiver can be found in~\cite{Taylor_11,Zhou_10,Liu_12},
where~\cite{Taylor_11,Zhou_10} considered the channel estimation error at the legitimate receiver and~\cite{Liu_12} considered the channel estimation error at both the legitimate receiver and the eavesdropper.

Specifically, Taylor et al. presented the impact of the legitimate receiver's channel estimation error on the performance of an eigenvector-based jamming technique in~\cite{Taylor_11}. Their research showed that the ergodic secrecy rate provided by the jamming technique decreases rapidly as the channel estimation error increases.
Zhou and McKay analyzed the optimal power allocation of the artificial noise for the secure transmission considering the impact of imperfect CSI at the legitimate receiver in~\cite{Zhou_10}. They found that it is wise to create more artificial noise by compromising on the transmit power of information-bearing signals when the CSI is imperfectly obtained.
Liu et al.~\cite{Liu_12} adopted the secrecy beamforming scheme to investigate the joint design of training and data transmission signals for wiretap channels.
They derived the ergodic secrecy rate for practical systems with imperfect channel estimations at both the legitimate receiver and the eavesdropper, and found the optimal tradeoff between the energy used for training and data signals based on the achievable ergodic secrecy rate.

The aforementioned works in~\cite{Taylor_11,Zhou_10,Liu_12} all used the ergodic secrecy rate to characterize the performance limits of systems.
The ergodic secrecy rate is an appropriate secrecy measure for systems in which the encoded messages span sufficient channel realizations to capture the ergodic features of the fading channel~\cite{Gopala_08}.
In addition, the works in~\cite{Taylor_11,Zhou_10,Liu_12} implicitly assumed variable-rate transmission strategies where the encoding rates are adaptively chosen according to the instantaneous channel gains\footnote{The system achieving the ergodic secrecy rate has the implicit assumption of the variable-rate transmission, which is very different from traditional ergodic fading scenarios without the secrecy consideration. A detailed explanation can be found in~\cite{Gopala_08}.}.
In practice, communication systems sometimes prefer non-adaptive rate transmission to reduce complexity and applications like video streams in multimedia often require fixed encoding rates\footnote{In this paper, systems with non-adaptive rates are different from systems with fixed rates.
The systems with fixed rates indicate that the encoding rates are already given and hence cannot be chosen freely.
The systems with non-adaptive rates indicate that the encoding rates can be chosen in the design process but are constant for all message transmissions.}.
Thus, variable-rate transmission strategies are not always feasible.

In this paper, we study the secure on-off transmission design with channel estimation errors, and adopt an outage-based characterization as the security  performance measurement. Here the secure on-off transmission scheme is adopted from~\cite{Gopala_08,Zhou_11} and is essential to control the secrecy performance for systems with fixed encoding rates.

The main contributions of this paper are summarized as follows.

\begin{enumerate}
  \item We consider quasi-static slow fading channels and use an outage-based formulation to study the secure transmission design with channel estimation errors at the receiver side. This is different from the previous works in~\cite{Taylor_11,Zhou_10,Liu_12} which used the ergodic secrecy rate as the performance measure.

  \item   We develop throughput-maximizing secure on-off transmission schemes with fixed encoding rates for different scenarios distinguished on whether or not there is channel estimation error at the eavesdropper, and whether or not the transmitter has the estimated channel quality fed back from the eavesdropper.
      Our analytical and numerical results show how the optimal design and the achievable throughput vary with the change in the channel knowledge assumptions.

  \item For systems in which the encoding rates are controllable parameters to design, we jointly optimize the encoding rates and the on-off transmission thresholds to maximize the throughput of secure transmissions. Both non-adaptive and adaptive rate transmissions are considered.
      Note that none of the previous works on physical layer security considering the channel estimation error has explicitly involved the rate parameters as part of the design problem.

  \item We also analyze how the training (pilot) power affects the achievable throughput of secure transmissions, since the accuracy of the channel estimation depends on the pilot power. One interesting finding is that, in the scenario where both the legitimate receiver and the eavesdropper obtain imperfect channel estimates, increasing the pilot power for more accurate channel estimation can harm the throughput of the secure transmission even if the pilot power is obtained for free.
\end{enumerate}

The remainder of this paper is organized as follows.
Section~\ref{sec:SysMod} gives the system model and the assumptions on channel knowledge.
Section~\ref{sec:Design1} analyzes the secure on-off transmission design for systems with fixed encoding rates.
Section~\ref{sec:Design2} develops two joint rate and on-off transmission designs depending on whether the encoding rates are non-adaptive or adaptive.
Numerical results and conclusions are given in Sections~\ref{sec:Numernew} and~\ref{sec:Conc}, respectively.

\section{System Model} \label{sec:SysMod}
We consider a wireless communication system in which the transmitter, Alice, wants to send confidential information to the intended user, Bob, in the presence of an eavesdropper, Eve.
Alice, Bob and Eve are assumed to have a single antenna each.
We consider the scenario where both Bob and Eve are mobile users served by the base station, Alice.
In order to secure the transmission to Bob against Eve, Alice tracks the channel qualities of both mobile stations by asking them to feed back their estimated instantaneous channel qualities through error-free feedback links\footnote{Note that only the channel quality, which is a real number as opposed to the complex channel coefficient, is required to be fed back to Alice. In this paper, we assume a high-quality feedback link with negligible quantization errors.}.

The main assumptions on the system model made in this paper are listed as follows.
\begin{enumerate}[(a)]
  \item We assume quasi-static fading channels and adopt the block fading model~\cite{Ozarow_94}, where the channel gains remain constant over a block of symbols (i.e., the transmission of one message) and change independently from one block to the next.
  \item The block-wise transmission is adopted. At the start of each block, pilot symbols are transmitted to enable channel estimation at the receiver. Then, both Bob and Eve estimate their channels and feed the estimated channel qualities back to Alice. Finally, the data symbols are transmitted.
  \item We assume that the transmission power of the pilot symbol can be different from the transmission power of the data symbol.
  \item We assume that the duration of a block is sufficiently long. For simplicity, the time spent on training and feedback is negligible compared with the data transmission time.
  \item We assume that the average signal-to-noise ratios (SNRs) at both Bob and Eve, without the consideration of channel estimation errors, are known at Alice.
\end{enumerate}
Among the above five assumptions, assumptions~(a), (b), (e) are more important than assumptions~(c) and~(d). With assumptions~(a) and~(b), the receivers are able to provide instantaneous CSI feedback to Alice. Assumption~(e) indicates that Alice can obtain the statistics of both Bob and Eve's channels.
Note that the availability of (some forms of) instantaneous and statistical CSI at the transmitter is very important for designing secure transmission schemes at the physical layer. Assumption (c) is very general and practical, which is often assumed in the work related to pilot-symbol-aided channel estimation.
Assumption (d) is made for the simplicity of presenting and discussing results.
This assumption can be relaxed if needed.
For example, if the block length is not sufficient long such that the time spent on training and feedback is considerable compared with the data transmission time, we can introduce a new parameter to represent the ratio of pilot transmission and feedback time to data transmission time, and include this parameter when evaluating the system performance.

The data symbol transmitted by Alice is denoted by $d$. The transmission power of the data symbol is normalized so that $E\{|d|^2\}=1$, where $E\{\cdot\}$ is the expectation operation. The pilot symbol is denoted by $t$.
The ratio of pilot power to data power is denoted by
\begin{equation}\label{}
  \alpha=\frac{E\{|t|^2\}}{E\{|d|^2\}}=E\{|t|^2\}.
\end{equation}
Since $E\{|d|^2\}=1$, we also call $\alpha$ as the normalized pilot power (normalized by data power) in this paper.
The received symbols at Bob and Eve are, respectively, given by
\begin{equation}\label{}
  y_b=\sqrt{P_b}h_bx+n_b,
\end{equation}
\begin{equation}\label{}
  y_e=\sqrt{P_e}h_ex+n_e,
\end{equation}
where $h_b$ and $h_e$ denote the channel gains from Alice to Bob and Alice to Eve, respectively.
each having a zero-mean complex Gaussian distribution with unit variance, i.e., $\mathcal{CN}(0,1)$.
We assume that Bob and Eve's channels, i.e., $h_b$ and $h_e$, are independent. This assumption is reasonable for rich-scattering environment where Bob and Eve are not very close to each other.
The additive white noise with complex Gaussian distribution $\mathcal{CN}(0,1)$ at Bob and Eve are denoted by $n_b$ and $n_e$. The transmitted signal $x$ can be a data symbol, $d$, or a pilot symbol, $t$. Since the data power is normalized to unity, $P_b$ and $P_e$ represent the average data signal-to-noise ratios at Bob and Eve without the consideration of channel estimation errors, respectively. Thus, $P_b$ and $P_e$ are parameters that indicate the general channel conditions between the transmitter and the receivers. For example, $P_b>P_e$ may indicate that the distance from Alice to Bob is smaller than the distance from Alice to Eve.

\subsection{Channel Estimation}\label{subsec:ChannelEstimation}
We assume that Bob's channel is estimated by the MMSE estimator during pilot transmission. The estimation of Bob's channel gain and the estimation error are denoted by $\hat{h}_b$ and $\tilde{h}_b$, respectively. Thus,
\begin{equation}\label{eq:ohtb}
  h_b=\hat{h}_b+\tilde{h}_b,
\end{equation}
where $\hat{h}_b$ and $\tilde{h}_b$ are assumed to have zero-mean complex Gaussian distributions.
The assumption of Gaussian distributed channel estimation error arises from using the MMSE estimator for channel estimation in the Bayesian linear model~\cite{Kay_93}, e.g., pilot-symbol-aided channel estimation~\cite{Cavers_91} as considered in this work. More specifically, since the channel coefficient, $h_b$, has a complex Gaussian distribution and the received signal, $y_b$, is a linear function of the channel coefficient, the linear MMSE estimation becomes the optimal MMSE estimation. Thus, by using a linear estimator, the estimated channel coefficient and the estimation error are zero-mean complex Gaussian distributed.
In fact, $|\hat{h}_b|$ is what Bob feeds back to Alice as the estimated instantaneous channel quality.
The orthogonality principle implies $E\{|h_b|^2\}=E\{|\hat{h}_b|^2\}+E\{|\tilde{h}_b|^2\}$.
According to~\cite{Hassibi_03}, the variance of channel estimation error is given by
\begin{equation}\label{eq:PtVsBetaB}
  \beta_b=E\{|\tilde{h}_b|^2\}=\frac{1}{1+\alpha P_bT_t},
\end{equation}
where $T_t$ is the length of pilot transmission. In this paper, it is assumed that $T_t=1$. Hence the effect of channel training is solely characterized by the normalized pilot power, $\alpha$.
For convenience, we let $\vrhb=P_b|\hat{h}_b|^2$ and $\vrtb=P_b|\tilde{h}_b|^2$, each having an exponential distribution given by
\begin{equation}\label{}
f_{\vrhb}(\vrhb)=\frac{1}{\vrhba}\exp\left(-\frac{\vrhb}{\vrhba}\right), ~~~~ \vrhb>0,
\end{equation}
\begin{equation}\label{}
f_{\vrtb}(\vrtb)=\frac{1}{\vrtba}\exp\left(-\frac{\vrtb}{\vrtba}\right), ~~~~ \vrtb>0.
\end{equation}
We assume that Bob uses the estimated channel gain for data detection. Then, the actual instantaneous SNR at Bob can be written as~\cite{Vakili_06}
\begin{equation}\label{eq:vr_b}
  \vr_b=\frac{P_b|\hat{h}_b|^2}{P_b|\tilde{h}_b|^2+1}=\frac{\vrhb}{\vrtb+1}.
\end{equation}

We assume that Eve's channel is also estimated by the MMSE estimator.
The estimation of Eve's channel gain and the estimation error are denoted by $\hat{h}_e$ and $\tilde{h}_e$, respectively. Thus,
\begin{equation}\label{eq:ohte}
  h_e=\hat{h}_e+\tilde{h}_e.
\end{equation}
Under the assumption of using MMSE estimator for channel estimation in the Bayesian linear model, $\hat{h}_e$ and $\tilde{h}_e$ have zero-mean complex Gaussian distributions.
In fact, $|\hat{h}_e|$ is what Eve is required to feed back to Alice as the estimated instantaneous channel quality.
The orthogonality principle implies $E\{|h_e|^2\}=E\{|\hat{h}_e|^2\}+E\{|\tilde{h}_e|^2\}$. In addition, the variance of channel estimation error is given by
\begin{equation}\label{eq:PtVsBetaE}
  \beta_e=E\{|\tilde{h}_e|^2\}=\frac{1}{1+\alpha P_eT_t},
\end{equation}
where we assume $T_t=1$.
Similar, we let $\vrhe=P_e|\hat{h}_e|^2$ and $\vrte=P_e|\tilde{h}_e|^2$, each having an exponential distribution given by
\begin{equation}\label{}
f_{\vrhe}(\vrhe)=\frac{1}{\vrhea}\exp\left(-\frac{\vrhe}{\vrhea}\right), ~~~ \vrhe>0,
\end{equation}
\begin{equation}\label{}
f_{\vrte}(\vrte)=\frac{1}{\vrtea}\exp\left(-\frac{\vrte}{\vrtea}\right), ~~~~ \vrte>0.
\end{equation}
With the MMSE channel estimation, the actual instantaneous SNR for data detection at Eve can be written as
\begin{equation}\label{eq:vr_e1}
  \vr_e=\frac{P_e|\hat{h}_e|^2}{P_e|\tilde{h}_e|^2+1}=\frac{\vrhe}{\vrte+1}.
\end{equation}

It should be noted that in principle Eve is able to further improve the channel estimation by performing joint channel and data detection, while Alice has no mechanism to tell if this is the case. As a robust approach for achieving secrecy, Alice may assume the worst case scenario where Eve perfectly knows her own channel. Then, the actual instantaneous SNR at Eve is $\vr_e=P_e|h_e|^2$, which has an exponential distribution given by
\begin{equation}\label{eq:vr_e2}
f_{\vr_e}(\vr_e)=\frac{1}{\vrea}\exp\left(-\frac{\vr_e}{\vrea}\right), ~~~~ \vr_e>0.
\end{equation}

\subsection{Channel Knowledge}\label{subsec:ChannelKnowledge}
As mentioned before, Alice asks both Bob and Eve to feed back their estimated instantaneous channel qualities after the pilot transmission phase.
Since Bob is the intended user, we simply assume that Alice has and trusts the feedback from Bob with the knowledge of $\vrhb=P_b|\hat{h}_b|^2$ as Bob's estimated instantaneous SNR. The actual instantaneous SNR at Bob is given in (\ref{eq:vr_b}).
However, Eve is an eavesdropper, and may not cooperate with Alice. Hence, Alice may not obtain or trust the feedback information from Eve.
In this work, we specifically investigate the following three scenarios with different assumptions on the channel knowledge: \vspace{1.5mm}
\begin{itemize}
  \item Scenario 1: Alice has and trusts the feedback from Eve, knowing $\vrhe=P_e|\hat{h}_e|^2$ as the estimate of the instantaneous SNR at Eve. Eve uses the MMSE channel estimate $\hat{h}_e$ for data detection, and hence the actual instantaneous SNR at Eve is given in (\ref{eq:vr_e1}).
  \item Scenario 2: Alice has and trusts the feedback from Eve, knowing $\vrhe=P_e|\hat{h}_e|^2$ as the estimate of the instantaneous SNR at Eve. Eve is assumed to perfectly know her own channel, and the actual instantaneous SNR at Eve is $\vr_e=P_e|h_e|^2$.
  \item Scenario 3: Alice does not have or trust Eve's feedback, and hence has no knowledge about Eve's instantaneous channel. However, the statistics of Eve's channel, i.e., $P_e$, is still assumed to be known at Alice.
      Eve perfectly knows her own channel, and the actual instantaneous SNR at Eve is $\vr_e=P_e|h_e|^2$.
\end{itemize}

In fact, the three scenarios above can also be interpreted as follows.
Scenario 1 represents the case where Eve is exactly identical to other mobile users.
Scenario 2 generally represents the case where Alice has partial information about Eve's channel gain, while allowing Eve to have perfect knowledge on her own channel.
Scenario 3 is valid for the case where Alice has no feedback from Eve. This scenario is perhaps the most practical one with current communication protocols where the channel feedback is only obtained from the intended receiver. Scenario 3 is also a robust approach for secrecy that allows Eve to have malicious behaviors, e.g., feeding wrong information back to Alice.

It should be noted that Scenario~2 is the least-practical scenario compared with Scenarios~1 and~3. However, it is necessary to stress the value of studying Scenario~2 in this paper.
From the legitimate users' perspectives, Scenario~1 represents the most desirable case, where Alice has the feedback from Eve and Eve has imperfect CSI.
In contrast, Scenario~3 represents the worst case, where Alice has no feedback from Eve and Eve has perfect CSI.
There are two different CSI assumptions between these two scenarios, one on the feedback from Eve to Alice and the other on the CSI knowledge at Eve.
From theoretical point of view, it is meaningful to see the impact of changing one of the CSI assumptions on the secure transmission design.
To this end, Scenario~2 is introduced as it only differs from Scenario~1 or~3 in one CSI assumption.
Thus, Scenario~2 enables us to compare the secure transmissions with different CSI assumptions changing in step.
For instance, the difference in the results between Scenarios~1 and~2 shows the effect of the CSI quality at Eve. The difference between Scenarios~2 and~3 shows the effect of the availability of CSI feedback.

\subsection{Secure Encoding}\label{subsec:SecureEncoding}
We consider the widely-adopted wiretap code~\cite{wyner_75} for confidential message transmissions. There are two rate parameters, namely, the codeword transmission rate, $R_b$, and the confidential information rate, $R_s$. The positive rate difference $R_e=R_b-R_s$ is the cost to provide secrecy against the eavesdropper. A length $M$ wiretap code is constructed by generating $2^{MR_b}$ codewords $x^M(w,v)$ of length $M$, where $w=1,2,\cdots,2^{MR_s}$ and $v=1,2,\cdots,2^{M(R_b-R_s)}$. For each message index $w$, we randomly select $v$ from $\left\{1,2,\cdots,2^{M(R_b-R_s)}\right\}$ with uniform probability and transmit the codeword $x^M(w,v)$.
From~\cite{wyner_75}~\cite[Theorem~1]{Thangaraj_07}~\cite[Definition~2]{Tang_09}, perfect secrecy cannot be achieved when $R_e<C_e$, where $C_e$ denotes Eve's channel capacity, $C_e=\log_2(1+\vr_e)$. Also, Bob is unable to decode the received codewords correctly when $R_b>C_b$, where $C_b$ denotes Bob's channel capacity, $C_b=\log_2(1+\vr_b)$.
Thus, given a pair of the rate choices, $R_b$ and $R_s$, the secrecy outage probability~\cite{Zhou_11}, $p_{so}$, and the connection outage probability, $p_{co}$, are defined as
\begin{equation}\label{eq:psoGen}
  p_{so}=\Pr(C_e>R_b-R_s \mid \text{message transmission}),
\end{equation}
\begin{equation}\label{eq:pcoGen}
  p_{co}=\Pr(C_b<R_b \mid \text{message transmission}),
\end{equation}
where $\Pr(\cdot)$ denotes the probability measure.
Note that both outage probabilities are conditioned on the message transmission.
The security level and the reliability level of a transmission scheme can then be measured by the secrecy outage probability and the connection outage probability, respectively.

\section{On-Off Transmission Design} \label{sec:Design1}
In this section, we consider each of the three scenarios described in Section~\ref{sec:SysMod} and show how to design transmission schemes with good throughput performance, whilst satisfying certain constraints on the reliability and security levels.
In particular, we consider the on-off transmission:
Alice decides whether or not to transmit according to the information about Bob and Eve's estimated instantaneous SNRs, i.e., transmission takes place when the estimated instantaneous SNR at Bob, $\vrhb$, is greater than a certain threshold, $\mu_b$, and the estimated instantaneous SNR at Eve, $\vrhe$, is less than another threshold, $\mu_e$, while transmission is suspended when $\vrhb\le\mu_b$ or $\vrhe\ge\mu_e$.
Having this on-off transmission scheme is necessary for improving the reliability and security performance.
For example, when the channel condition from Alice to Bob is very bad, transmission may incur a large probability of decoding error at Bob. Also, when the channel condition from Alice to Eve is very good, transmitting message may lead to a large probability that the confidential information is leaked to Eve.
Since the security and reliability performances are related to different channels, which can be seen from (\ref{eq:psoGen}) and~(\ref{eq:pcoGen}), it is reasonable to set two separate SNR thresholds on Bob's channel and Eve's channel, respectively.
In the scenario where Alice does not have or trust the feedback from Eve, there is no on-off SNR threshold on Eve's channel, $\mu_e$, or equivalently $\mu_e=\infty$.

We assume that the encoding rates have already been designed such that both the codeword transmission rate, $R_b$, and the confidential information rate, $R_s$, are fixed\footnote{The problem considering the design of encoding rates where $R_b$ and $R_s$ can be optimally chosen is analyzed in Section~\ref{sec:Design2}.}.
The design problem is to maximize the throughput, $\eta$, subject to two constraints, one on the security performance and the other on the reliability performance, which can be written as \footnote{Note that we do not consider the overhead of pilot and feedback when calculating the throughput in this paper, since we assume a sufficiently long block length for simplicity. If the pilot transmission and feedback time is considered, we can introduce a new parameter, say $\theta$, to represent the ratio of pilot transmission and feedback time to data transmission time. Then, the throughput can be calculated by taking this ratio, $\theta$, into account, i.e., (\ref{eq:eta111}) will change to
$ \eta=\frac{1}{1+\theta}p_{tx}(1-p_{co})R_s.$
}
\begin{eqnarray}\label{eq:optprobgiven}
  \max_{\mu_b, \mu_e} &&  \eta=p_{tx}\left(1-p_{co}\right)R_s, \label{eq:eta111} \\
  \text{s.t.}&& p_{so}\le\epsilon, p_{co}\le\delta, \label{eq:constrainsinonofffum}
\end{eqnarray}
where $p_{tx}$ denotes the probability of transmission due to the on-off transmission scheme, $\epsilon\in [0,1]$ and $\delta\in [0,1]$ represent the security and reliability requirements. The secrecy outage probability is required to be no larger than $\epsilon$, and the connection outage probability is required to be no larger than $\delta$.
The controllable parameters to design are the two on-off SNR thresholds, $\mu_b$ and $\mu_e$.
Note that the throughput maximization provided in this paper only gives an achievable
bound on the throughput of secure transmission.

In what follows, we consider the transmission design in the three different scenarios described in Section \ref{sec:SysMod}. For each scenario, the transmission probability, the connection outage probability and the secrecy outage probability are derived firstly. Then, the feasibility of security and reliability constraints is discussed. Here the feasibility of constraints means that the constraints can be satisfied whilst achieving a positive information rate. Finally, the solution of the optimization problem is given as a proposition.
\subsection{Scenario One}\label{subsec:Design1_S1}
\vspace{1mm}{\em{\underline{Derivations of $p_{tx}, p_{co}$ and $p_{so}$:}}}
Since Bob's estimated instantaneous SNR is independent with Eve's estimated instantaneous SNR, the probability of transmission in Scenario~1 is given as
\begin{eqnarray}\label{eq:ptx1}
\!\!\!\!\!\!p_{tx}\!\!\!&=&\!\!\!\Pr(\vrhb>\mu_b)\Pr(\vrhe<\mu_e)  \nonumber \\
\!\!\!&=&\!\!\!\exp\!\left(\!-\frac{\mu_b}{\vrhba}\right)\!\!\left(1-\exp\!\left(\!-\frac{\mu_e}{\vrhea}\right)\!\right).~
\end{eqnarray}
Since $\vr_b\le\vrhb$ according to (\ref{eq:vr_b}) and Bob can decode the message without error only when $C_b\ge R_b$, it is wise to choose the value of $\mu_b$ satisfying
\begin{equation}\label{eq:Non_LowerBound_mu_o}
  \log_2(1+\mu_b)\ge R_b \Rightarrow \mu_b\ge2^{R_b}-1.
\end{equation}
Then, the connection outage probability in Scenario~1 is given by
\begin{eqnarray}\label{eq:pco1}
p_{co}\!\!\!&=&\!\!\!\Pr\left(\log_2(1+\vr_b)<R_b\mid\vrhb>\mu_b\right) \nonumber\\
\!\!\!&=&\!\!\!\Pr\left(\log_2\left(1+\frac{\vrhb}{\vrtb+1}\right)<R_b\mid\vrhb>\mu_b\right) \nonumber\\
\!\!\!&=&\!\!\!\frac{\Pr(\mu_b<\vrhb<(2^{R_b}-1)(\vrtb+1))}{\Pr(\vrhb>\mu_b)}\nonumber\\
\!\!\!&=&\!\!\!\exp\left(\frac{\mu_b}{\vrhba}\right)\nonumber\\
\!\!\!&&\!\!\!\cdot\int_{\frac{\mu_b}{2^{R_b}-1}\!-\!1}^{\infty}\!\left(\!\int_{\mu_b}^{(2^{R_b}-1)(\vrtb+1)}\!\!\!\!{f_{\vrhb}(\vrhb)\mathrm{d}{\vrhb}}\!\right)\!\!f_{\vrtb}(\vrtb)\mathrm{d}{\vrtb} \nonumber \\
\!\!\!&=&\!\!\!\frac{\beta_b(2^{R_b}-1)}{1+\beta_b(2^{R_b}-2)}\exp\left(\!\frac{1}{\vrtba}\left(\!1-\frac{\mu_b}{2^{R_b}-1}\!\right)\!\right).
\end{eqnarray}
The secrecy outage probability in Scenario~1 is given by
\begin{eqnarray}\label{eq:pso1_1}
  p_{so}&=&\Pr(C_e>R_b-R_s\mid\vrhe<\mu_e)   \nonumber \\
  &=&\Pr\left(\log_2\left(1+\frac{\vrhe}{\vrte+1}\right)>R_b-R_s\mid\vrhe<\mu_e\right) \nonumber\\
  &=&\frac{\Pr\left((2^{R_b-R_s}-1)(\vrte+1)<\vrhe<\mu_e\right)}{\Pr(\vrhe<\mu_e)}.
\end{eqnarray}
On one hand, if $\mu_e\le2^{R_b-R_s}-1$, $p_{so}=0$. On the other hand, if $\mu_e>2^{R_b-R_s}-1$, we have
\begin{eqnarray}\label{eq:pso1_2}
p_{so}\!\!\!\!\!&=&\!\!\!\!\!\frac{\int_0^{\frac{\mu_e}{2^{R_{b}-R_{s}}-1}-1}
\!\!\!\left(\!\int_{(2^{R_b-R_s}-1)(\vrte+1)}^{\mu_e}\!f_{\vrhe}\!(\vrhe)\mathrm{d}{\vrhe}\!\right)
\!f_{\vrte}\!(\vrte)\mathrm{d}{\vrte}}
{1-\exp\left(-\frac{\mu_e}{\vrhea}\right)} \nonumber \\
\!\!\!\!\!&=&\!\!\!\!\!\frac{\frac{1-\beta_e}{1+\beta_e(2^{R_b-R_s}-2)}\exp\!\left(\!-\frac{2^{R_b-R_s}-1}{\vrhea}\right)\!-\!\exp\left(\!-\frac{\mu_e}{\vrhea}\right)}{1-\exp\left(-\frac{\mu_e}{\vrhea}\right)}      \nonumber\\
\!\!\!\!\!&&\!\!\!\!\!+\frac{\frac{\beta_e(2^{R_b-R_s}-1)}{1+\beta_e(2^{R_b-R_s}-2)}\exp\!\left(\!\frac{1}{P_e}\!\left(\!\frac{1}{\beta_e}\!-\!\frac{\mu_e}{1-\beta_e}\!-\!\frac{\mu_e}{\beta_e(2^{R_b-R_s}-1)}\!\right)\!\right)}{1-\exp\left(-\frac{\mu_e}{\vrhea}\right)}.\nonumber \\
\end{eqnarray}
From (\ref{eq:pso1_1}) and (\ref{eq:pso1_2}), we find that the secrecy outage probability is directly influenced by the value of $\mu_e$ but not related to $\mu_b$. If $\mu_e\le2^{R_b-R_s}-1$, perfect secrecy is achievable in Scenario~1. Since $\vrhe\ge\vr_e$ in Scenario~1, the estimate of Eve's instantaneous SNR, in fact, can be treated as an upper bound of the actual Eve's instantaneous SNR. Hence, Alice can make sure $C_e<R_b-R_s$ as long as $\mu_e\le2^{R_b-R_s}-1$, and then the perfect secrecy is achieved. If $\mu_e>2^{R_b-R_s}-1$, (\ref{eq:pso1_2}) indicates that the secrecy outage probability increases as the value of $\mu_e$ increases.

\vspace{1mm}{\em{\underline{Feasibility of Constraints:}}}
From (\ref{eq:pco1}), $p_{co}$ is a decreasing function of $\mu_b$ and
\begin{equation}\label{eq:limpcoS1}
  \lim_{\mu_b\rightarrow+\infty}p_{co}=0.
\end{equation}
Thus, the feasible range of the reliability constraint in Scenario~1 is given by
\begin{equation}\label{eq:fea_delta_1}
  0<\delta\le1.
\end{equation}
According to (\ref{eq:pso1_1}), $p_{so}$ is an increasing function of $\mu_e$ and $p_{so}=0$ as long as $\mu_e\le2^{R_b-R_s}-1$. Thus, the feasible range of the security constraint in Scenario~1 is given by
\begin{equation}\label{}
  0\le\epsilon\le1.
\end{equation}
Hence, any required reliability and security constraints are feasible by appropriately adjusting the on-off thresholds. It is noted that perfect secrecy, i.e., $\epsilon=0$,  can be achieved.

The following proposition summarizes the solution to the design problem in Scenario~1, where the optimal $\mu_b$ is expressed in a closed form and the optimal $\mu_e$ is obtained by numerically solving an equation.

\begin{Proposition}\label{Pro:S1}
{\em{The optimal parameters of the throughput-maximizing transmission scheme in Scenario~1 are given as follows:
\begin{flalign}\label{eq:mu_bopt1}
\mu_b=\begin{cases}
  2^{R_b}-1, ~\quad \quad \quad \quad \quad \text{if} \quad R_b\le\log_2\left(1+\frac{(1-\beta_b)\delta}{\beta_b(1-\delta)}\right),\\
  \!\left(2^{R_b}-1\right)\!\left(1-{\vrtba}\ln\left(\delta\frac{1+\beta_b(2^{R_b}-2)}{\beta_b(2^{R_b}-1)}\!\right)\right),~\text{otherwise.}
  \end{cases}
\end{flalign}
\begin{flalign}\label{eq:mu_eopt1}
\mu_e=\begin{cases}
  +\infty , \quad &\text{if} ~~ \frac{1-\beta_e}{1+\beta_e(2^{R_b-R_s}-2)}\exp\left(-\frac{2^{R_b-R_s}-1}{\vrhea}\right)\le\epsilon,\\
  F_1,  &\text{otherwise,}
  \end{cases}
\end{flalign}
\\where $F_1$ is the solution of $\mu_e$ to the equation
\begin{eqnarray}\label{eq:mu_eopt1fun}
\epsilon\!\!\!\!\!&=&\!\!\!\!\!\frac{\frac{1-\beta_e}{1+\beta_e(2^{R_b-R_s}-2)}\exp\!\left(\!-\frac{2^{R_b-R_s}-1}{\vrhea}\right)\!-\!\exp\left(\!-\frac{\mu_e}{\vrhea}\right)}{1-\exp\left(-\frac{\mu_e}{\vrhea}\right)}      \nonumber\\
\!\!\!\!\!&&\!\!\!\!\!+\frac{\frac{\beta_e(2^{R_b-R_s}-1)}{1+\beta_e(2^{R_b-R_s}-2)}\exp\!\left(\!\frac{1}{P_e}\!\left(\!\frac{1}{\beta_e}\!-\!\frac{\mu_e}{1-\beta_e}\!-\!\frac{\mu_e}{\beta_e(2^{R_b-R_s}-1)}\!\right)\!\right)}{1-\exp\left(-\frac{\mu_e}{\vrhea}\right)}.\nonumber\\
\end{eqnarray}
}}
\end{Proposition}
The proof of this proposition is given in Appendix~\ref{App:ProofPro1}.

\vspace{1mm} {{\em{Remark:}}
From (\ref{eq:pco1}), when the reliability constraint is very stringent such that $p_{co}$ is required to go to zero, the value of the on-off SNR threshold on Bob's channel needs to be very large such that $\mu_b$ goes to infinity. However, if $\mu_b$ goes to infinity, we have the throughput, $\eta$, goes to zero. Thus, it is interesting to investigate the behaviors of $\eta$ and $p_{co}$ for the limiting case where $\mu_b$ goes to infinity\footnote{ Note that $\eta$ also goes to zero, as $\mu_e$ goes to zero. However, since perfect secrecy is achievable as long as $\mu_e\le2^{R_b-R_s}-1$ in this scenario, it is unnecessary to study the behavior of $\eta$ as $\mu_e$ goes to zero.}. From (\ref{eq:eta111}), (\ref{eq:ptx1}) and (\ref{eq:pco1}), we see that both $\eta$ and $p_{co}$ are exponential functions of $\mu_b$ as $\mu_b$ goes to infinity. Hence, the slopes of $\eta$ and $p_{co}$, as a function of $\mu_b$, both go to zero as $\mu_b$ goes to infinity.

In addition, in this scenario if the transmitter increases the pilot power, the estimation errors at both the legitimate receiver and the eavesdropper will reduce. Thus, the selection of normalized pilot power, $\alpha$, will create an interesting tradeoff between reducing the estimation errors at the legitimate receiver and reducing the estimation errors at the eavesdropper. Here, we briefly discuss the method to calculate the optimal $\alpha$ as follows, instead of providing a detailed analysis. First, we need to find the expressions of optimal $\mu_b$ and $\mu_e$ in terms of $\alpha$ by substituting (\ref{eq:PtVsBetaB}) and~(\ref{eq:PtVsBetaE}) into (\ref{eq:mu_bopt1}) and~\ref{eq:mu_eopt1}), respectively. Then, $p_{tx}$ and $p_{co}$ can be expressed as functions of $\alpha$. Finally, the optimal $\alpha$ is the solution to the optimization problem of
\begin{eqnarray}
 \max_{\alpha}&& \eta=p_{tx}(\alpha)\left(1-p_{co}(\alpha)\right)R_s, \\
  s.t.&& \alpha>0.
\end{eqnarray}
Due to the complex expressions of the optimal $\mu_b$ and $\mu_e$, it is difficult to find a closed-form solution of the optimal $\alpha$. But this problem can be solved numerically.

\subsection{Scenario Two}\label{subsec:Design1_S2}
\vspace{1mm}{\em{\underline{Derivations  of $p_{tx}, p_{co}$ and $p_{so}$:}}}
The derivations  of the probability of transmission and the connection outage probability in Scenario~2 are the same as (\ref{eq:ptx1}) and (\ref{eq:pco1}) in Scenario~1, respectively.
The secrecy outage probability in Scenario~2 is given by
\begin{eqnarray}\label{eq:pso2}
  p_{so}\!\!\!&=&\!\!\!\Pr(C_e>R_b-R_s\mid\vrhe<\mu_e)   \nonumber \\
  \!\!\!&=&\!\!\!\Pr(\log_2(1+\vr_e)>R_b-R_s\mid\vrhe<\mu_e) \nonumber \\
  \!\!\!&=&\!\!\!\frac{\Pr(\vr_e>2^{R_b-R_s}-1, \vrhe<\mu_e)}{\Pr(\vrhe<\mu_e)}  \nonumber \\
  \!\!\!&=&\!\!\!\frac{\int_0^{\mu_e}\left(\int_{2^{R_b-R_s}-1}^\infty f_{\vr_e|\vrhe}(\vr_e|\vrhe)\mathrm{d}{\vr_e}\right)f_{\vrhe}(\vrhe)\mathrm{d}{\vrhe}}{1-\exp\left(-\frac{\mu_e}{\vrhea}\right)}.~~~~
\end{eqnarray}
According to the definitions of $\vr_e$ and $\vrhe$ in Scenario~2, $\vr_e$ conditioned on its estimate, $\vrhe$, follows a non-central chi-square distribution with two degrees of freedom. Applying the cumulative distribution function of the non-central chi-square distribution, we have
\begin{equation}\label{}
  \int_{2^{R_b-R_s}\!-\!1}^\infty \!\!f_{\vr_e|\vrhe}(\vr_e|\vrhe)\mathrm{d}{\vr_e}\!=\!Q_1\!\left(\!\sqrt{\frac{2\vrhe}{\vrtea}}, \!\sqrt{\frac{2^{R_b-R_s+1}\!-\!2}{\vrtea}}\right),~~
\end{equation}
where $Q_x(a,b)$ represents the Marcum Q-function~\cite{Marcum_50}.
Thus, the secrecy outage probability in Scenario~2 can be rewritten as
\begin{eqnarray}\label{eq:pso2_2}
\!\!\!\!\!\!\!\!p_{so}\!\!\!\!\!&=&\!\!\!\!\!\frac{\int_0^{\mu_e}\!Q_1\!\left(\!\sqrt{\!\frac{2\vrhe}{\vrtea}}, \sqrt{\!\frac{2^{R_b\!-\!R_s\!+\!1}-2}{\vrtea}}\right)\!f_{\vrhe}\!(\vrhe)\mathrm{d}{\vrhe}}{1\!-\!\exp\!\left(\!-\frac{\mu_e}{\vrhea}\!\right)}\nonumber\\
\!\!\!\!\!&=&\!\!\!\!\!\frac{\int_0^{\mu_e}\!\!\exp\!\left(\!-\frac{\vrhe}{\vrhea}\!\right)\!Q_1\!\left(\!\sqrt{\!\frac{2\vrhe}{\vrtea}}, \sqrt{\!\frac{2^{R_b\!-\!R_s\!+\!1}-2}{\vrtea}}\right)\!\mathrm{d}{\vrhe}}
{P_e\left(1\!-\!\beta_e\right)
\left(\!1\!-\!\exp\!\left(\!-\frac{\mu_e}{\vrhea}\!\right)\!\right)}.~~
\end{eqnarray}

\vspace{1mm}{\em{\underline{Feasibility of Constraints:}}}
Since the connection outage probability does not change from Scenario~1 to Scenario~2, the feasible range of the reliability constraint in Scenario~2 is identical to (\ref{eq:fea_delta_1}) in Scenario~1.
Since $p_{so}$ is an increasing function of $\mu_e$ and
\begin{eqnarray}\label{}\label{eq:psolimitation}
  \lim_{\mu_e\rightarrow0}p_{so}&=& \Pr(C_e>R_b-R_s\mid\vrhe=0) \nonumber \\
  &=&\Pr(\log_2(1+\vr_e)>R_b-R_s\mid\vrhe=0) \nonumber \\
  &=&\int_{2^{R_b-R_s}-1}^\infty f_{\vr_e|\vrhe=0}(\vr_e|\vrhe=0)\mathrm{d}{\vr_e}  \nonumber \\
  &=&Q_1\left(0, \sqrt{\frac{2^{R_b-R_s+1}-2}{\vrtea}}\right).
\end{eqnarray}
Thus, the feasible range of the security constraint is given as
\begin{equation}\label{eq:feasible_epsilon_s2}
  Q_1\left(0, \sqrt{\frac{2^{R_b-R_s+1}-2}{\vrtea}}\right)<\epsilon\le1.
\end{equation}
Thus, any required reliability constraint is feasible, while the security constraint is feasible only when (\ref{eq:feasible_epsilon_s2}) is satisfied.

The following proposition summarizes the solution to the design problem in Scenario~2, where the optimal $\mu_b$ is expressed in a closed form and the optimal $\mu_e$ is obtained by numerically solving an equation.

\begin{Proposition}\label{Pro:S2}
{\em{The optimal parameters of the throughput-maximizing transmission scheme in Scenario~2 are given as follows:
\begin{flalign}\label{eq:mu_bopt2}
\mu_b=\begin{cases}
  2^{R_b}-1, ~\quad \quad \quad \quad \quad ~\text{if} \quad R_b\le\log_2\left(1+\frac{(1-\beta_b)\delta}{\beta_b(1-\delta)}\right),\\
  \!\left(2^{R_b}-1\right)\!\left(1-{\vrtba}\ln\left(\delta\frac{1+\beta_b(2^{R_b}-2)}{\beta_b(2^{R_b}-1)}\!\right)\right),~\text{otherwise.}
  \end{cases}
\end{flalign}
\begin{flalign}\label{eq:mu_eopt2}
\!\!\!\!\!\!\!\!\!\!\!\!\!\!\!\!\!\!\!\!\!\!\!\!\!\!\!\!\!\!\!\!\!\!\!\!\!\!\!\!\mu_e=\begin{cases}
  +\infty ,   \quad &\text{if} ~~
  \exp\left(-\frac{2^{R_b-R_s}-1}{\vrea}\right)\le\epsilon,\\
  F_2,   &\text{otherwise,}
  \end{cases}
\end{flalign}
\\where $F_2$ is the solution of $\mu_e$ to the equation
\begin{equation}\label{eq:F2SolutioninS2}
\epsilon\!=\!\frac{\int_0^{\mu_e}\!\!\exp\!\left(\!-\frac{\vrhe}{\vrhea}\!\right)\!Q_1\!\left(\!\sqrt{\!\frac{2\vrhe}{\vrtea}}, \sqrt{\!\frac{2^{R_b\!-\!R_s\!+\!1}-2}{\vrtea}}\right)\!\mathrm{d}{\vrhe}}
{P_e\left(1\!-\!\beta_e\right)
\left(\!1\!-\!\exp\!\left(\!-\frac{\mu_e}{\vrhea}\!\right)\!\right)}.
\end{equation}
}}
\end{Proposition}
The proof of this proposition is given in Appendix~\ref{App:ProofPro2}.
Note that the optimal $\mu_b$ in Scenario~2 is identical to that in Scenario~1.

\vspace{1mm} {\em{Remark:}}
In this scenario, when the security constraint is very stringent such that $p_{so}$ converges to its limit in \eqref{eq:psolimitation}, the value of the on-off SNR threshold on Eve's channel needs to be very small such that $\mu_e$ goes to zero. However, if $\mu_e$ goes to zero, we have the throughput, $\eta$, goes to zero. Thus, it is interesting to investigate the behavior of $\eta$ for the limiting case where $\mu_e$ goes to zero or equivalently $p_{so}$ converges to its limit\footnote{The behavior of $\eta$ for the limiting case where $\mu_b$ goes to infinite in this scenario is exactly the same as discussed in Scenario~1. To avoid the redundancy, we do not discuss it here again.}
From (\ref{eq:eta111}) and~(\ref{eq:ptx1}), $\eta$ can be rewritten as
\begin{equation}
   \eta(\mu_e) = A\left(1-\exp(-B\mu_e)\right),
\end{equation}
where $A=\exp\left(-\frac{\mu_b}{\vrhba}\right)(1-p_{co})R_s$ and $B=\frac{1}{P_e(1-\beta_e)}$.
The Taylor expansion of the above function around $\mu_e=0$ is given by
\begin{eqnarray}
\sum_{n=0}^{\infty}\frac{\eta^{(n)}(0)\mu_e^n}{n!}&=& A\left(1-\sum_{n=0}^{\infty}(-1)^n\frac{B^n\mu_e^n}{n!}\right) \nonumber \\
  &=&A\left(1-\left(1-B\mu_e+O\left(\mu_e^2\right)\right)\right) \nonumber \\
  &=&AB\mu_e-O\left(\mu_e^2\right),
\end{eqnarray}
where $O(\cdot)$ denotes the less-significant terms, and expresses the error.
Thus, the most-significant term of $\eta(\mu_e)$ around $\mu_e=0$ is
\begin{equation}
  AB\mu_e=\frac{(1-p_{co})R_s}{P_e(1-\beta_e)}\exp\left(-\frac{\mu_b}{\vrhba}\right)\mu_e,
\end{equation}
and the slope of $\eta(\mu_e)$, as $\mu_e$ goes to zero, can be approximated as
 \begin{equation}
   \frac{(1-p_{co})R_s}{P_e(1-\beta_e)}\exp\left(-\frac{\mu_b}{\vrhba}\right).
 \end{equation}

Besides, according to (\ref{eq:mu_eopt2}) in Proposition~\ref{Pro:S2}, $\mu_e=\infty$ when
\begin{equation}\label{eq:loose_epsilon_S2}
  \exp\left(-\frac{2^{R_b-R_s}-1}{\vrea}\right)\le\epsilon\le1.
\end{equation}
This indicates that Alice can ignore the feedback from Eve to design the system parameters when the security constraint satisfies (\ref{eq:loose_epsilon_S2}). Therefore, the design problem in Scenario~2 is identical to the design problem in Scenario~3 when the security constraint satisfies (\ref{eq:loose_epsilon_S2}).

\subsection{Scenario Three}\label{subsec:Design1_S3}
In Scenario~3, Alice does not have or trust the feedback from Eve. Thus, Alice decides whether or not to transmit according to the information about Bob's estimated instantaneous SNR. Then, the on-off SNR threshold on Eve's channel, $\mu_e$, does not exist, and there is only one parameter to design, i.e., $\mu_b$.

\vspace{1mm}{\em{\underline{Derivations of $p_{tx}, p_{co}$ and $p_{so}$:}}}
The probability of transmission in Scenario~3 is given as
\begin{equation}\label{eq:ptx3}
  p_{tx}=\Pr(\vrhb>\mu_b)=\exp\left(-\frac{\mu_b}{\vrhba}\right).
\end{equation}
The derivation of the connection outage probability in Scenario~3 is identical to (\ref{eq:pco1}) in Scenarios~1 and 2.
The secrecy outage probability in Scenario~3 is given by
\begin{equation}\label{eq:pso3}
  p_{so}=\Pr(C_e>R_b-R_s)=\exp\left(-\frac{2^{R_b-R_s}-1}{\vrea}\right).
\end{equation}
Note that the secrecy outage probability in Scenario~3 is a constant value and uncontrollable. Thus, the security constraint is either always achievable or always unachievable no matter what the value of the design parameter is.

\vspace{1mm}{\em{\underline{Feasibility of Constraints:}}}
Since the connection outage probability remains the same in Scenarios 1, 2 and 3, the feasible range of the reliability constraint in Scenario~3 is identical to (\ref{eq:fea_delta_1}) in Scenarios~1 and 2.
Since the secrecy outage probability in Scenario~3 is not controllable, the feasible range of the security constraint in Scenario 3 is given by
\begin{equation}\label{eq:feasible_epsilon_s3}
  \exp\left(-\frac{2^{R_b-R_s}-1}{\vrea}\right)\le\epsilon\le1.
\end{equation}
Thus, any required reliability constraint is feasible, while the security constraint is feasible only when (\ref{eq:feasible_epsilon_s3}) is satisfied.
Note that the lower bound of the feasible security constraint in this scenario is the same as (\ref{eq:loose_epsilon_S2}) in the analysis for Scenario~2.
This is because the design problems in Scenarios~2 and 3 are the same when (\ref{eq:loose_epsilon_S2}) is satisfied.

The following proposition summarizes the solution to the design problem in Scenario~3.
\begin{Proposition}\label{Pro:S3}
{\em{The optimal parameter of the throughput-maximizing transmission scheme in Scenario~3 is given in (\ref{eq:mu_bopt1}).
}}
\end{Proposition}

\vspace{1mm}{\em{Remark:}}
Comparing the optimal solutions to the design problems in the three different scenarios,
we can find that the three scenarios have the same optimal solution of $\mu_b$ but different optimal solutions of $\mu_e$. This is because that we have the same assumption on the channel knowledge of the legitimate link but different assumptions on the channel knowledge of the eavesdropper's link in different scenarios.

Besides, it is noted that the security performance of systems in Scenario 3 cannot be controlled by the design parameters for the fixed rate transmission scheme. In order to control the security performance of systems in Scenario~3, a detailed analysis on the joint rate and on-off transmission design for systems in Scenario~3 is provided in the next section.

\section{Joint Rate and On-Off Transmission Design} \label{sec:Design2}
As analyzed in Section~\ref{sec:Design1}, for networks in Scenario~3, the security performance of the communication system is uncontrollable if we only consider the design of the on-off transmission parameters, i.e, the on-off thresholds.
In order to control the security performance, in this section, we re-study the design problem in Scenario~3 considering the joint rate and on-off transmission design\footnote{The joint rate and on-off transmission design for Scenarios~1 and 2 can be obtained in a similar way as presented in this section.}.
Unlike the on-off transmission design in Section~\ref{sec:Design1} where the encoding rates, $R_b$ and $R_s$, are fixed, in this section we allow more degrees of freedom  such that $R_b$ and $R_s$ can be optimally chosen.

The design problem is to maximize the throughput, $\eta$, subject to two constraints, one on the security performance and the other on the reliability performance. In Scenario~3, Alice decides whether or not to transmit according to the estimated instantaneous SNR at Bob, $\vrhb$.
The design problem can be written as
\begin{eqnarray}\label{}
  \max_{\mu_b, R_b, R_s} && \eta,\\
  \text{s.t.}&& p_{so}\le\epsilon, p_{co}\le\delta.
\end{eqnarray}
The controllable parameters to design are the codeword transmission rate, $R_b$, the confidential information rate, $R_s$, and the on-off SNR threshold on Bob's channel, $\mu_b$. In the following, two different transmission schemes are derived, according to whether the encoding rates are non-adaptive or adaptive.
The expression of the throughput, $\eta$, for each transmission scheme is provided in the corresponding subsection.

\subsection{Non-Adaptive Rate Scheme}\label{sec:subNonAda}
We first consider the non-adaptive rate scheme where the codeword transmission rate, $R_b$, and the confidential information rate, $R_s$, are both constant over time.
The throughput for the non-adaptive rate scheme is given by
\begin{equation}\label{}
  \eta=p_{tx}(1-p_{co})R_s.
\end{equation}

\vspace{1mm}{\em{\underline{Derivations of $p_{tx}, p_{co}$ and $p_{so}$:}}}
The probability of transmission is given in (\ref{eq:ptx3}). The connection outage probability is given in (\ref{eq:pco1}). The secrecy outage probability is given in (\ref{eq:pso3}).
Note that the security performance is controllable now, since $R_b$ and $R_s$ can be optimal chosen.

\vspace{1mm}{\em{\underline{Feasibility of Constraints:}}}
Since $p_{so}$ is independent of $\mu_b$, the choice of $\mu_b$ does not affect $p_{so}$. Also, from (\ref{eq:limpcoS1}), we can set $\mu_b$ sufficiently large to achieve any arbitrarily small $p_{co}$.
Thus, the feasible range of the reliability constraint in the non-adaptive rate scheme is identical to (\ref{eq:fea_delta_1}).
According to (\ref{eq:pso3}), $p_{so}$ is a decreasing function of $R_b-R_s$ and
\begin{equation}\label{}
  \lim_{R_b-R_s\rightarrow+\infty}p_{so}=0.
\end{equation}
Thus, the feasible range of the security constraint in the non-adaptive rate scheme is given by
\begin{equation}\label{eq:fea_eps_fix_overall}
  0<\epsilon\le1.
\end{equation}
Note that any required reliability and security constraints are feasible by appropriately choosing $R_b$ and $R_s$.

In Section~\ref{sec:Design1}, $p_{so}$ and $p_{co}$ are independently controlled by different design parameters.
However, in this section, the choices of encoding rates affect both the connection outage probability and the secrecy outage probability.
In other words, with the encoding rates controllable, $p_{so}$ and $p_{co}$ are related by the rate parameters.
For example, from the derivations of connection and secrecy outage probabilities, a smaller $R_b$ allows us to achieve a smaller connection outage probability but may increase the  secrecy outage probability.
This enables a trade-off between the feasible reliability constraint and the feasible security constraint.
To illustrate such a trade-off, we analyze the feasible constraints for the system with a given on-off threshold, $\mu_b$.
To satisfy $R_s>0$ and $p_{so}\le\epsilon$, we have $2^{R_b}-1>\vrea\ln\epsilon^{-1}.$
Also, from (\ref{eq:Non_LowerBound_mu_o}) and $p_{co}\le\delta$, we have
$2^{R_b}-1\le\min\left\{\mu_b, F_4(\mu_b, \delta)\right\}$
where $F_4(\mu_b, \delta)$ is the positive solution of $x$ to the equation
\begin{equation}\label{eq:FFF}
  \mu_b= x\left(1-{\vrtba}\ln\left(\delta\frac{\beta_bx+1-\beta_b}{\beta_bx}\right)\right).
\end{equation}
Thus, for any chosen value of $\mu_b$, the feasible constraints for having secure communication with positive confidential information rate must satisfy
\begin{equation}\label{eq:Non_constraints}
  \exp\left(-\frac{\min\left\{\mu_b, F_4(\mu_b, \delta)\right\}}{\vrea}\right)<\epsilon.
\end{equation}
From (\ref{eq:FFF}), it is easy to see that $F_4(\mu_b, \delta)$ is an increasing function of $\delta$. Thus, according to (\ref{eq:Non_constraints}), the minimum feasible value of $\epsilon$ increases with the decrease of $\delta$. In other words, if we set a stricter reliability constraint, the feasible security constraint becomes loose.
Note that when the reliability constraint is sufficiently loose, $F_4(\mu_b, \delta)$ becomes always greater than $\mu_b$, and (\ref{eq:Non_constraints}) changes to
\begin{equation}\label{eq:Non_constraints_lower}
  \exp\left(-\frac{\mu_b}{\vrea}\right)<\epsilon.
\end{equation}

The following proposition summarizes the solution to the design problem for the non-adaptive rate scheme, where each of the optimal $\mu_b$ and the optimal $R_s$ is expressed as a closed-form function of $R_b$ and the optimal $R_b$ is obtained by numerically solving an optimization problem.
\begin{Proposition}\label{Pro:Non_Ada}
{\em{The optimal parameters of the throughput-maximizing transmission scheme with non-adaptive rates are given as follows:
\begin{flalign}\label{eq:mu_bopt_fix}
\mu_b=\begin{cases}
  2^{R_b}-1, ~\quad \quad \quad \quad \quad ~\text{if} \quad R_b\le\log_2\left(1+\frac{(1-\beta_b)\delta}{\beta_b(1-\delta)}\right),\\
  \!\left(2^{R_b}-1\right)\!\left(1-{\vrtba}\ln\left(\delta\frac{1+\beta_b(2^{R_b}-2)}{\beta_b(2^{R_b}-1)}\!\right)\right),~\text{otherwise.}
  \end{cases}
\end{flalign}
\begin{equation}\label{eq:Non_Rs}
\!R_s=R_b-k,   \qquad \text{where} \quad k=\log_2(1+\vrea\ln\epsilon^{-1}).
\end{equation}
\!$R_b$ is obtained by solving the problem given as
\begin{eqnarray}\label{eq:Non_Rb}
\!\!\!\max_{R_b}\!\!\!\!&&\!\!\!\!
(R_b-k)
\exp\left(-\frac{\mu_b}{\vrhba}\right)\nonumber\\
\!\!\!\!&&\!\!\!\!\cdot\left(\!1\!-\!\frac{\beta_b\!\left(2^{R_b}\!-\!1\right)}{1\!+\!\beta_b\!\left(2^{R_b}\!-\!2\right)}
\exp\!\left(\!\frac{1}{\vrtba}\!\left(\!1\!-\!\frac{\mu_b}{2^{R_b}\!-\!1}\!\right)\!\right)\!\right),~~~~\\
\text{s.t.}\!\!\!\!&&\!\!\!\!k<R_b<\nonumber\\
&&\!\!\!\!\!\!\max\!\left\{\!\log_2\!\!\left(\!\!1\!\!+\!\!\frac{(1\!-\!\beta_b)\delta}{\beta_b(1\!-\!\delta)}\!\right)\!\!, k\!+\!\!\frac{1}{\ln\!2}\!\mathrm{W}\!\!\left(2^{-k}P_b(1\!-\!\beta_b)\right)\!\!\right\}\!,
\end{eqnarray}
where $\mathrm{W}(\cdot)$ is the Lambert W function and $\mu_b$ is a function of $R_b$ whose expression is formulated as (\ref{eq:mu_bopt_fix}). }}
\end{Proposition}
The proof of this proposition is given in Appendix~\ref{App:ProofPro4}

\subsection{Adaptive Rate Scheme}\label{subsec:Ada}
Now, we consider the scenario where the codeword transmission rate, $R_b$, and the confidential information rate, $R_s$, can be adaptively chosen according to the estimated Bob's instantaneous SNR.
Since the confidential information rate, $R_s$, is adaptively chosen according to any given $\vrhb$, the throughput for the adaptive rate scheme is given by
\begin{equation}\label{eq:th_ada}
  \eta=\int_{\mu_b}^\infty{(1-p_{co})R_s}f_{\vrhb}(\vrhb)\mathrm{d}{\vrhb}.
\end{equation}
The lower limit of the integral in (\ref{eq:th_ada}) is equal to $\mu_b$, since the transmission takes place only when $\vrhb>\mu_b$ due to the on-off transmission scheme.

Then, we consider the design problem of finding the values of $R_b, R_s$ and $\mu_b$ that maximize the throughput. Since $R_b$ and $R_s$ can be adaptively chosen according to any given $\vrhb$, we treat this design as a two-step optimization problem given by
\vspace{1mm}\\
Step~1: For any given $\vrhb$ ($\vrhb>\mu_b$), solve
\begin{eqnarray}\label{}
\max_{R_b, R_s}\!\!&&\!\!\left(1-p_{co}\right)R_s,  \\
\text{s.t.}\!\!&&\!\!p_{so}\le\epsilon, p_{co}\le\delta.
\end{eqnarray}
Step~2: Choose the best $\mu_b$ to maximize the overall throughput averaged over $\vrhb$.\\
Note that the optimal $R_b$ and $R_s$ are obtained in Step~1 for a given value of $\vrhb$. Thus, the following calculations of connection and secrecy outage probabilities are conditioned on a given $\vrhb$.

\vspace{1mm}{\em{\underline{Derivations of $p_{co}$ and $p_{so}$:}}}
Since $\vr_b\le\vrhb$ and Bob can decode the message without error only when $C_b\ge R_b$, it is wise to choose the value of $R_b$ satisfying $R_b\le\log_2(1+\vrhb)$.
Then, for any given $\vrhb$, the connection outage probability can be computed as
\begin{eqnarray}\label{eq:Ada_pco}
  p_{co}&=&\Pr\left(\log_2(1+\vr_b)<R_b \mid \vrhb\right) \nonumber\\
  &=&\Pr\left(\log_2\left(1+\frac{\vrhb}{\vrtb+1}\right)<R_b \mid \vrhb\right) \nonumber\\
  &=&\Pr\left(\vrtb>\frac{\vrhb}{2^{R_b}-1}-1 \mid \vrhb\right) \nonumber\\
  &=&\exp\left(-\frac{1}{\vrtba}\left(\frac{\vrhb}{2^{R_b}-1}-1\right)\right).
\end{eqnarray}
The secrecy outage probability does not change from the non-adaptive rate scheme given in (\ref{eq:pso3}).

\vspace{1mm}{\em{\underline{Feasibility of Constraints:}}}
According to (\ref{eq:Ada_pco}), we have
\begin{equation}\label{}
  \vrhb\rightarrow\infty \Rightarrow p_{co}\rightarrow0
\end{equation}
Since $p_{so}$ is independent of $\mu_b$, the choice of $\mu_b$ does not affect $p_{so}$. Also, we can set $\mu_b$ sufficiently large such that transmission happens only when $\vrhb$ is sufficiently large to achieve any arbitrarily small $p_{co}$. Therefore, it is feasible to have $\delta\rightarrow0$.
Thus, the feasible range of the reliability constraint is the same as (\ref{eq:fea_delta_1}).
For the same reason described in the non-adaptive rate scheme, the feasible range of the security constraint is identical to (\ref{eq:fea_eps_fix_overall}).
Therefore, any required reliability and security constraints are feasible by appropriately choosing $R_b$ and $R_s$.

The following proposition summarizes the solution to the design problem for the adaptive rate scheme, where the optimal $\mu_b$ is given by a closed-form solution, the optimal $R_s$ is expressed as a closed-form function of $R_b$ and the optimal $R_b$ is obtained by numerically solving an optimization problem.
\begin{Proposition}\label{Pro:Ada}
{\em{ The optimal parameters of the throughput-maximizing transmission scheme with adaptive rates are given as follows:
\begin{equation}\label{eq:Ada_mu}
\!\!\!\!\!\!\!\!\!\!\!\!\!\!\!\!\!\!\!\!\!\!\!\!\!\!\!\!\!\!\!\!\!\!\!\!\!\!\!\!\!\!\!\!\!\!\!\mu_b=\left(1+\vrtba\ln\delta^{-1}\right)\vrea\ln\epsilon^{-1}.
\end{equation}
\begin{equation}\label{eq:Ada_Rs}
R_s=R_b-k,   \quad \text{where} ~~ k=\log_2(1+\vrea\ln\epsilon^{-1}).
\end{equation}
$R_b$ is obtained by solving the problem given by
\begin{eqnarray}
\max_{R_b}&&\!\!\!\!\!\! (R_b\!-\!k)\left(1\!-\!\exp\left(\frac{1}{\vrtba}\left(1\!-\!\frac{\vrhb}{2^{R_b}\!-\!1}\right)\right)\right), \\
\text{s.t.}&&\!\!\!\!\!\!k<R_b\le\log_2\left(1\!+\!\frac{\vrhb}{1+\vrtba\ln\delta^{-1}}\right).
\end{eqnarray}
}}
\end{Proposition}
The proof of this proposition is given in Appendix~\ref{App:ProofPro5}.

\vspace{1mm} {\em{Remark:}}
From Proposition~\ref{Pro:Ada}, one can further obtain that the optimal $R_b$ is equal to either the upper bound of $R_b$, i.e., $R_b=\log_2\left(1+\frac{\vrhb}{1+\vrtba\ln\delta^{-1}}\right)$, or the solution of $R_b$ to the equation
\begin{eqnarray}\label{}
  \frac{\mathrm{d}I(R_b)}{\mathrm{d}R_b}=0
\end{eqnarray}
where $I(R_b)=(R_b-k)\left(1-\exp\left(\frac{1}{\vrtba}\left(1-\frac{\vrhb}{2^{R_b}-1}\right)\right)\right)$.
Note that when $\beta_b=0$, Proposition~\ref{Pro:Ada} implies that $R_b=\log_2(1+\vr_b)$. This is consistent with the optimal solution of $R_b$ in the absence of the estimation error, where the optimal codeword rate matches the capacity of Bob's channel.

\section{Numerical Results}\label{sec:Numernew}
In this section, we illustrate and analyze the numerical results for both the on-off transmission design and the joint rate and on-off transmission design.
\subsection{On-off Transmission Design}\label{subsec:Numerical_D1}
We first present and compare the numerical results for the on-off transmission designs in the three different scenarios.
The results shown in this subsection are all for networks with the transmission rates fixed to $R_b=2$ and $R_s=1$.

\begin{figure}[!htb]
\centering\vspace{-0mm}
\includegraphics[width=1\columnwidth]{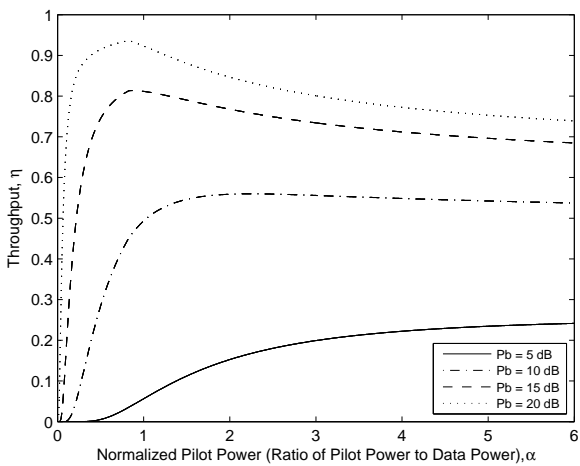}
\vspace{-6mm} \caption{Scenario~1: Achievable throughput versus normalized pilot power. Results are shown for networks with different average received data SNRs at Bob, $P_b =$ 5 dB, 10 dB, 15 dB, 20 dB. The other system parameters are $\delta=0.1$, $\epsilon=0.05$, $P_e=0$ dB,  $R_b=2$, $R_s=1$.} \label{fig:S1_1}
\end{figure}

\begin{figure}[!htb]
\centering\vspace{-0mm}
\includegraphics[width=1\columnwidth]{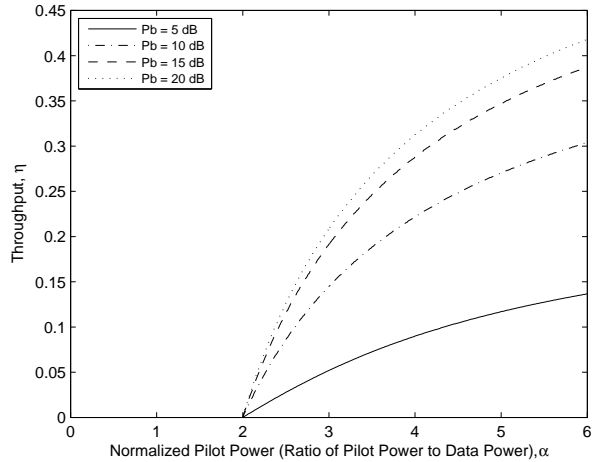}
\vspace{-6mm} \caption{Scenario~2: Achievable throughput versus normalized pilot power. Results are shown for networks with different average received data SNRs at Bob, $P_b =$ 5 dB, 10 dB, 15 dB, 20 dB. The other system parameters are $\delta=0.1$, $\epsilon=0.05$, $P_e=0$ dB,  $R_b=2$, $R_s=1$.} \label{fig:S2_1}
\end{figure}

Figs.~\ref{fig:S1_1} and ~\ref{fig:S2_1} demonstrate the achievable throughput against the normalized pilot power for networks with different average received SNRs at Bob\footnote{The results with $P_b$ equal to or smaller than $P_e$ are not shown in the figures. When $P_b$ is comparable or small than $P_e$, the achievable throughput is very small or reaches zero. In order to achieve better performance in such a scenario, one can consider multi-antenna transmissions or using external helpers to regain the relative advantage of the legitimate receiver's channel over the eavesdropper's channel, which is beyond the scope of this work.} in Scenarios~1 and 2, respectively. The average received SNR at Eve, $P_e$, is fixed to 0 dB. Also, the reliability and  security constraints are fixed.
As shown in Fig.~\ref{fig:S1_1}, it is interesting that the throughput does not always increase with the increase of normalized pilot power. As the curves of $P_b =$ 10 dB, 15 dB, 20 dB present, the throughput increases fast to a peak when the normalized pilot power increases to the optimal value ($\alpha=2.28$ for $P_b = 10$  dB, $\alpha=0.87$ for $P_b = 15$ dB, $\alpha=0.83$ for $P_b = 20$ dB). After achieving the peak value, the throughput decreases with the increase of the normalized pilot power.
This observation can be explained as follows.
In scenario~1, both Bob and Eve estimate their channels via the pilot transmission and feed the channel estimates back to Alice.
Increasing pilot power not only enhances the legitimate users' knowledge about the channels, which has a positive effect on the secure transmission, but also increases the accuracy of channel estimation at the eavesdropper, which incurs a negative effect on the secure transmission.
Before the normalized pilot power reaches the optimal value, to obtain good channel knowledge at the legitimate users is more important than to keep the eavesdropper's channel estimation inaccurate.
However, after the pilot power reaches the optimal value, the disadvantage incurred by further increasing pilot power overcomes the benefit. Thus, we observe that increasing normalized pilot power incurs the peak value of throughput.
This observation suggests that when both Bob and Eve have imperfect channel estimation dependent on the training process, it is not always good to have more training power to get more accurate channel estimation, and the optimal value can be calculated according to the method discussed in the Remark of Section~\ref{subsec:Design1_S1}.

On the other hand, the achievable throughput always is a non-decreasing function of the normalized pilot power in Scenario~2, as shown in Fig.~\ref{fig:S2_1}.
In Scenario~2, only Bob has channel estimation errors but not Eve. Thus,
the increase of training power only improves the legitimate users' knowledge about the channels, but has no influence on the eavesdropper's knowledge about her own channel. Therefore, it is always good to have more training power to increase the throughput in this scenario.

\begin{figure}[!htb]
\centering\vspace{-0mm}
\includegraphics[width=1\columnwidth]{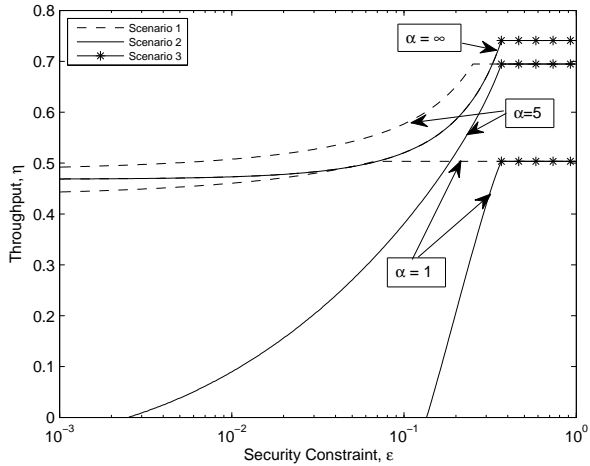}
\vspace{-6mm} \caption{Comparison of the three scenarios: Achievable throughput versus security constraint. Results are shown for networks with different values of normalized pilot power, $\alpha = 1, 5, \infty$.
The other system parameters are $P_b=10$ dB, $P_e=0$ dB, $\delta=0.1$, $R_b=2$, $R_s=1$.
Note that the case of $\alpha=\infty$ is equivalent to having perfect channel estimation.}
\label{fig:S1S2_1}
\end{figure}

Fig.~\ref{fig:S1S2_1} compares the achievable throughput in Scenarios~1, 2 and 3. There are three groups of curves representing the networks with three different values of normalized pilot power.
As shown in the figure, subject to different security constraints, Scenario~1 can always achieve a positive throughput. This is because Alice and Eve have the same amount of knowledge about the eavesdropper's channel in Scenario 1, and Alice in fact knows an upper bound of the actual instantaneous SNR at Eve ($\vrhe\ge\vr_e$).
On the other hand, Scenarios~2 and 3 can obtain a positive throughput only when the security constraints are in the feasible ranges as formulated in (\ref{eq:feasible_epsilon_s2}) and (\ref{eq:feasible_epsilon_s3}), respectively.
In addition, we see that the throughput of each network in Scenario~3 is a step function of the security constraint (the throughput is equal to either zero or a positive constant value), which is because that the controllable parameter is not related to the security performance of networks in Scenario~3.
Comparing the results for different scenarios, we see that the networks in the three scenarios can achieve the same throughput, when the security constraint is sufficiently loose satisfying (\ref{eq:loose_epsilon_S2}) or (\ref{eq:feasible_epsilon_s3}).
Besides, under a same security constraint, the throughput difference between networks in Scenarios~1 and 2 decreases with the increase of normalized pilot power.
As presented by the case of $\alpha=\infty$, Scenarios 1 and 2 can achieve the same throughput when the channel is perfectly estimated.

\begin{figure}[!htb]
\centering\vspace{-0mm}
\includegraphics[width=1\columnwidth]{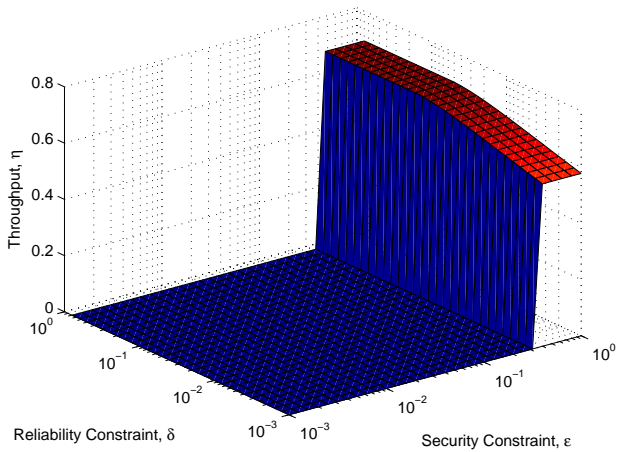}
\vspace{-6mm}   \caption{Scenario~3: Achievable throughput versus the security and reliability constraints. The system parameters are $\alpha=5$, $P_b=10$ dB, $P_e=0$ dB, $R_b=2$, $R_s=1$.}\label{fig:3D_GivenS3}
\end{figure}

Fig.~\ref{fig:3D_GivenS3} presents the achievable throughput versus the security and reliable constraints for the network in Scenario~3.
As shown in the figure, for different reliability constraints, the throughput is always a step function of the security constraint. Also, we find that the throughput increases with the loose of reliability  constraint at the beginning. However, if the reliability constraint is already sufficiently loose, further loosing the reliability constraint would not increase the throughput.

\subsection{Joint Rate and On-off Transmission Design}\label{subsec:Numerical_D2}
Now, we show the numerical results for the joint rate and on-off transmission design.
The results demonstrated in this subsection are obtained with $P_b=10$ dB and $P_e=0$ dB for networks in Scenario~3.

\begin{figure}[!h]
\centering\vspace{-0mm}
\includegraphics[width=1\columnwidth]{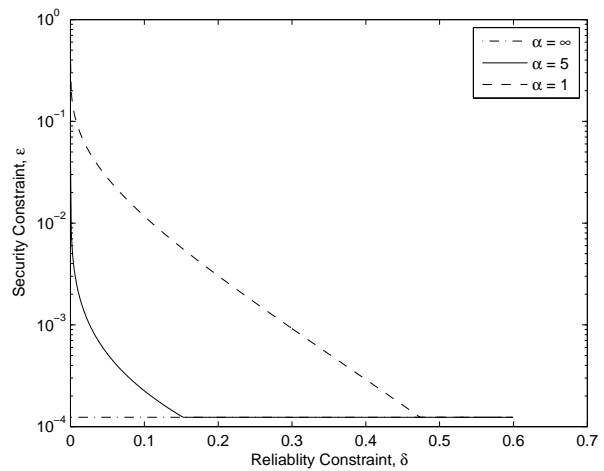}
\vspace{-6mm} \caption{Feasible security constraint versus feasible reliability constraint for non-adaptive rate scheme with a given on-off threshold. Results are shown for networks with different values of normalized pilot power, i.e., $\alpha=1, 5, \infty$. The other system parameters are $\mu_b=9$, $P_b=10$ dB, $P_e=0$ dB. } \label{fig:NonAdaptive_EpVsDe}
\end{figure}

Fig.~\ref{fig:NonAdaptive_EpVsDe} illustrates the trade-off between the feasible reliability constraint and the feasible security  constraint for the non-adaptive rate scheme with a given on-off threshold.
For each network, the feasible constraints lie in the region above the corresponding curve.
As depicted in the figure, there exists a lower bound on the feasible value of $\epsilon$, although the feasible value of $\epsilon$ generally decreases with the loose of reliability constraint.
From the analytical result, we know that the lower bound on the feasible value of $\epsilon$ is related to the on-off SNR threshold as given in (\ref{eq:Non_constraints_lower}).

\begin{figure}[!htb]
\centering\vspace{-0mm}
\includegraphics[width=1\columnwidth]{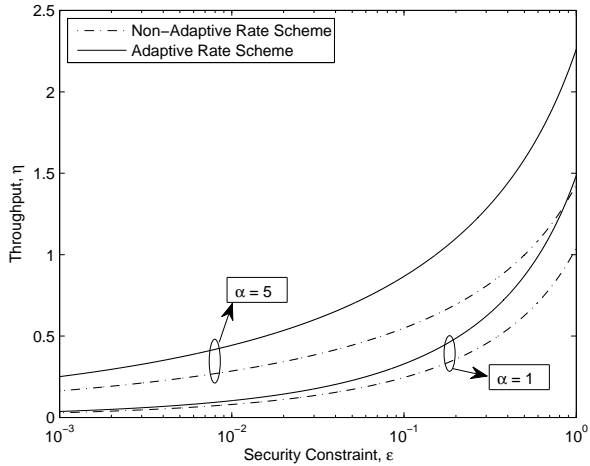}
\vspace{-6mm} \caption{Achievable throughput versus security constraint. Results are shown for networks with different values of normalized pilot power, i.e., $\alpha=1, 5$. The other system parameters are $\delta=0.1$, $P_b=10$ dB, $P_e=0$ dB.} \label{fig:ThVsEpsilon}
\end{figure}

Fig.~\ref{fig:ThVsEpsilon} demonstrates the achievable throughput over a range of security constraints for networks with different normalized pilot power values, while the reliability constraint is fixed to $\delta=0.1$.
The curves representing non-adaptive and adaptive rate schemes are distinguished by different line styles.
As shown in the figure, the achievable throughput rises with the increase of the normalized pilot power.
We see that  adaptively changing the encoding rates significantly improves the achievable throughput compared with the non-adaptive rate scheme.
In addition, compared with the on-off transmission design with fixed rates in Section~\ref{sec:Design1}, the joint rate and on-off transmission design significantly improves the achievable throughput.
For example, the on-off transmission design with fixed $R_b=2$ and $R_s=1$ cannot achieve a positive throughput value subject to a large range of security constraints, as shown in Fig.~\ref{fig:S1S2_1}, while the joint rate and on-off transmission design can always achieve a positive throughput value subject to any security constraint.

\begin{figure}[!htb]
\centering\vspace{-2mm}
\includegraphics[width=1\columnwidth]{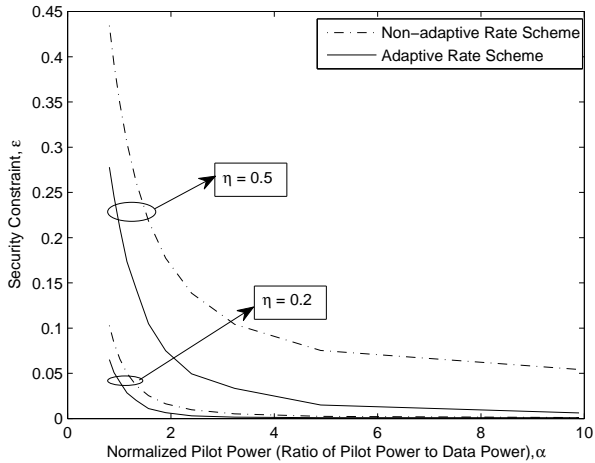}
\vspace{-6mm} \caption{Achievable security constraint versus normalized pilot power. Results are shown for networks with different target throughput values, i.e., $\eta=0.2, 0.5$. The other system parameters are $\delta=0.1$, $P_b=10$ dB, $P_e=0$ dB.} \label{fig:PilotVsEpsilon}
\end{figure}

Fig.~\ref{fig:PilotVsEpsilon} shows the effect of increasing the normalized pilot power on the achievable security level of networks with different target throughput values.
The curves representing non-adaptive and adaptive rate schemes are distinguished by different line styles.
By observing the slopes of curves, we find that the improvement of increasing the pilot power on the achievable security level is significant when the normalized pilot power is small. However, further increasing the pilot power can obtain very little benefit when the pilot power has already become large.

\begin{figure}[!htb]
\centering\vspace{-0mm}
\includegraphics[width=1\columnwidth]{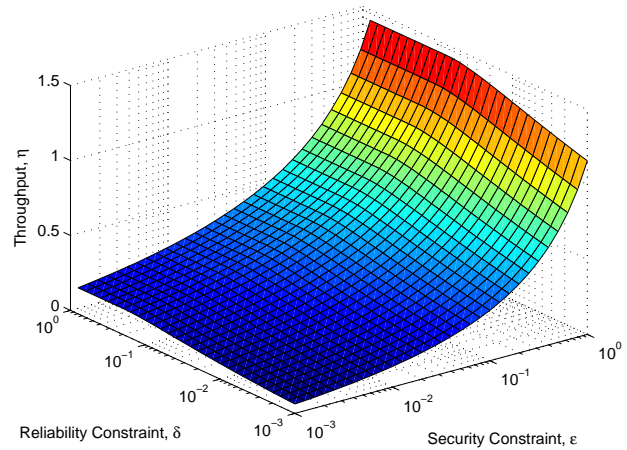}
\vspace{-6mm}  \caption{Non-adaptive rate scheme: Achievable throughput versus security and reliability constraints. The system parameters are $\alpha=5$, $P_b=10$ dB, $P_e=0$ dB.} \label{fig:3D_Fix}
\end{figure}

Fig.~\ref{fig:3D_Fix} presents the achievable throughput versus the security and reliable constraints for the non-adaptive rate scheme.
As shown in the figure, the throughput increases with the loose of security constraint all the time subject to a given reliability constraint. However, similar to the 3D result shown in the last subsection, the throughput increases only at the beginning with the loose of reliability  constraint subject to a given security constraint, and it would not continue increasing if the reliability constraint is already sufficiently loose.
In addition, we see that the change of the security constraint has larger effect on the throughput than the change of the reliability constraint, since the throughput increases faster along with the loose of security constraint than the loose of reliability constraint.

\section{Conclusions}\label{sec:Conc}
In this work, we presented a comprehensive study of secure transmission design in quasi-static slow fading channels with channel estimation errors. For systems with fixed encoding rates, throughput-maximizing on-off transmission schemes were proposed for scenarios with different assumptions on the channel knowledge. For systems with encoding rates controllable, we derived both non-adaptive and adaptive rate transmission schemes which jointly optimize the rate parameters and the on-off thresholds. Our analytical and numerical results illustrated how the optimal design and the achievable throughput vary with the change in the channel knowledge assumptions.
In addition, we found that increasing the pilot power for more accurate channel estimation sometimes can harm the system performance. When both the legitimate receiver and the eavesdropper estimate their channels via the pilot transmission,  increasing pilot power decreases the channel estimation errors at both the legitimate receiver and the eavesdropper. The overall throughput increases at the beginning but can decrease after achieving the peak value, as the pilot power increases.

Besides, some interesting research directions for the future work are discussed as follows.
In this paper, our design solutions are based on the assumption of independent channels for Bob and Eve. One interesting future work is to study the effect of channel correlation on the transmission design performance.
Also, apart from the channel estimation errors due to the estimation process, the investigation on the effects of additional channel estimation errors due to the quantization and finite-rate feedback is another interesting research direction.

\appendices
\section{}\label{App:ProofPro1}
\vspace{1mm} {\em{Proof of Proposition~\ref{Pro:S1}:}}
We first derive the optimal $\mu_b$ in Scenario~1.
One can find that $\mu_b=2^{R_b}-1$ is the only solution of $\mu_b$ to the equation
\begin{equation}\label{}
  \frac{\partial{\eta(\mu_b, \mu_e)}}{\partial{\mu_b}}=0
\end{equation}
and
\begin{equation}\label{}
  \frac{\partial^2\eta(2^{R_b}-1, \mu_e)}{\partial{\mu_b^2}}<0.
\end{equation}
Thus, if we ignore the possible bound of $\mu_b$, the optimal $\mu_b$ is equal to $2^{R_b}-1$.
However, to satisfy the reliability constraint, $p_{co}\le\delta$, there exists a possible lower bound of $\mu_b$ given by
\begin{equation}\label{eq:Non_LowerBound_mu}
  \mu_b\ge \left(2^{R_b}-1\right)\left(1-{\vrtba}\ln\left(\delta\frac{1+\beta_b(2^{R_b}-2)}{\beta_b(2^{R_b}-1)}\!\right)\right).
\end{equation}
Considering the lower bound, the optimal $\mu_b$ in Scenario~1 is formulated as (\ref{eq:mu_bopt1}) in Proposition~\ref{Pro:S1}.

Then, we derive the optimal $\mu_e$ in Scenario~1.
Since $p_{tx}$ is an increasing function of $\mu_e$ and $p_{co}$ is independent of $\mu_e$, it is optimal to maximize $\mu_e$ while satisfying the security constraint $p_{so}\le\epsilon$.
From the definition of $p_{so}$, one can find that $p_{so}$ is an increasing function of $\mu_e$. Thus, there is only one or no solution of $\mu_e$ to the equation
\begin{equation}\label{eq:solution_mu_e_S1}
  p_{so}(\mu_e)=\epsilon
\end{equation}
where the expression of $p_{so}$ is given as (\ref{eq:pso1_2}).
When
\begin{equation}\label{}
  \Pr(C_e>R_b-R_s)\le\epsilon \nonumber\\
\end{equation}
\begin{equation}
  \Leftrightarrow \frac{1-\beta_e}{1+\beta_e(2^{R_b-R_s}-2)}\exp\left(-\frac{2^{R_b-R_s}-1}{\vrhea}\right)\le\epsilon,
\end{equation}
there is no solution of $\mu_e$ to (\ref{eq:solution_mu_e_S1}), which means that there is no need to set an on-off SNR threshold on $\vrhe$ for the system (the required security constraint is always achievable) or equivalently $\mu_e=\infty$.
Otherwise, there exists one and only one solution of $\mu_e$ to (\ref{eq:solution_mu_e_S1}), which is the optimal value of $\mu_e$ to the maximization problem.
Although it is difficult to obtain a closed-form solution of $\mu_e$, this problem can be easily solved numerically.
Thus, the optimal $\mu_e$ in Scenario~1 is formulated as (\ref{eq:mu_eopt1}) in Proposition~\ref{Pro:S1}. \hfill$\blacksquare$

\section{}\label{App:ProofPro2}
\vspace{1mm} {\em{Proof of Proposition~\ref{Pro:S2}:}}
The optimal $\mu_b$ in Scenario~2 is the same as that in Scenario~1 and the proof of it is identical to the corresponding part in the proof of Proposition~\ref{Pro:S1}. Now, we derive the optimal $\mu_e$ in Scenario~2. Since $p_{tx}$ is an increasing function of $\mu_e$ and $p_{co}$ is independent of $\mu_e$, it is optimal to maximize $\mu_e$ while satisfying the security constraint $p_{so}\le\epsilon$.
From the definition of $p_{so}$, one can find that $p_{so}$ is an increasing function of $\mu_e$. Thus, there is only one or no solution of $\mu_e$ to the equation
\begin{equation}\label{eq:solution_mu_e_S2}
  p_{so}(\mu_e)=\epsilon
\end{equation}
where the expression of $p_{so}$ is given as (\ref{eq:pso2_2}).
When
\begin{equation}\label{}
  \Pr(C_e>R_b-R_s)\le\epsilon \Leftrightarrow \exp\left(-\frac{2^{R_b-R_s}-1}{\vrea}\right)\le\epsilon,
\end{equation}
there is no solution of $\mu_e$ to (\ref{eq:solution_mu_e_S2}), which means that there is no need to set an on-off SNR threshold on $\vrhe$ for the system (the required security constraint is always achievable) or equivalently $\mu_e=\infty$.
Otherwise, there exists one and only one solution of $\mu_e$ to (\ref{eq:solution_mu_e_S2}), which is the optimal value of $\mu_e$ to the maximization problem.
Although it is difficult to obtain a closed-form solution of $\mu_e$, this problem can be easily solved numerically.
Therefore, the optimal $\mu_e$ in Scenario~2 is formulated as (\ref{eq:mu_eopt2}) in Proposition~\ref{Pro:S2}. \hfill$\blacksquare$

\section{}\label{App:ProofPro4}
\vspace{1mm} {\em{Proof of Proposition~\ref{Pro:Non_Ada}:}}
The proof of the optimal $\mu_b$ for the non-adaptive scheme is identical to the proof of optimal $\mu_b$ in Section~\ref{sec:Design1}. Now, we prove the optimal $R_s$ for any chosen $R_b$ as follows. Since $p_{tx}$ and $p_{co}$ are independent of $R_s$, it is optimal to maximize $R_s$. Thus, we obtain the optimal $R_s$ while satisfying $p_{so}\le\epsilon$ as (\ref{eq:Non_Rs}) in Proposition~\ref{Pro:Non_Ada}.
Then, we prove the optimal $R_b$. Since $R_s>0$, we have $R_b>k$. It is easy to prove that when
\begin{equation}\label{}
R_b\ge\!\max\!\left\{\!\log_2\!\left(\!1\!+\!\frac{(1-\beta_b)\delta}{\beta_b(1-\delta)}\right)\!, k\!+\!\frac{1}{\ln\!2}\mathrm{W}\!\left(2^{-k}\vrhba\right)\!\right\}\!,~~
\end{equation}
the value of $\eta$ is a decreasing function of $R_b$, i.e,
\begin{equation}\label{}
  \frac{\partial{\eta(\mu_b, R_b)}}{\partial{R_b}}<0.
\end{equation}
Therefore, the optimal $R_b$ can be obtained by solving the optimization problem given in Proposition~\ref{Pro:Non_Ada}.\hfill$\blacksquare$

\section{}\label{App:ProofPro5}
\vspace{1mm} {\em{Proof of Proposition~\ref{Pro:Ada}:}}
The proof of the optimal $R_s$ for the adaptive rate scheme is identical to the corresponding part in the proof of Proposition~\ref{Pro:Non_Ada}.
Now, we derive the optimal $R_b$. To satisfy $R_s>0$ and $p_{co}\le\delta$, we obtain the lower and upper bounds of $R_b$ given by $R_b>k$ and $R_b\le\log_2\left(1+\frac{\vrhb}{1+\vrtba\ln\delta^{-1}}\right)$. Thus, the optimal $R_b$ can be obtained by solving the optimization problem given in Proposition~\ref{Pro:Ada}.
Then, we derive the optimal $\mu_b$. To derive the optimal, $\mu_b$, we start from looking for the range of $\vrhb$ in which it is possible to have secure communication with positive confidential information rate while satisfying both constraints. Let the lower bound of $R_b$ be less than the upper bound of $R_b$, we can find the feasible range of $\vrhb$ as
\begin{equation}\label{}
  \log_2\left(1+\vrea\ln\epsilon^{-1}\right)<\log_2\left(1+\frac{\vrhb}{1+\vrtba\ln\delta^{-1}}\right) \nonumber
\end{equation}
\begin{equation}\label{}
  \Leftrightarrow \vrhb>\left(1+\vrtba\ln\delta^{-1}\right)\vrea\ln\epsilon^{-1}.
\end{equation}
Therefore, the optimal $\mu_b$ is equal to the lower bound of the feasible $\vrhb$, given by (\ref{eq:Ada_mu}).\hfill$\blacksquare$

\bibliographystyle{IEEEtran}

\end{document}